\documentclass[%
 aip,
 rsi,
 reprint,%
 superscriptaddress,%
 amsmath,amssymb, %
floatfix,%
]{revtex4-2}

\usepackage{graphicx}
\usepackage{dcolumn}
\usepackage{bm}
\usepackage{siunitx}
\usepackage[dvipsnames]{xcolor}
\usepackage[normalem]{ulem}

\begin{document}

\preprint{AIP}

\title{Implementation of synthetic fast-ion loss detector and imaging heavy ion beam probe diagnostics in the 3D hybrid kinetic-MHD code MEGA}

\author{P. Oyola}
\email{poyola@us.es}
\affiliation{Department of Atomic, Molecular and Nuclear Physics, Universidad de Sevilla, Sevilla, 41012, Spain}

\author{J. Gonzalez-Martin}
\affiliation{Centro Nacional de Aceleradores (CNA) CSIC, 41092 Seville, Spain}
\affiliation{Department of Mechanical Engineering and Manufacturing, Universidad de Sevilla, Sevilla, 41092, Spain}

\author{M. Garcia-Munoz}
\affiliation{Department of Atomic, Molecular and Nuclear Physics, Universidad de Sevilla, Sevilla, 41012, Spain}
\affiliation{Centro Nacional de Aceleradores (CNA) CSIC, 41092 Seville, Spain}

\author{J. Galdon-Quiroga}
\affiliation{Max Planck Institute for Plasma Physics, 85748 Garching, Germany}
\author{G. Birkenmeier}
\affiliation{Max Planck Institute for Plasma Physics, 85748 Garching, Germany}
\affiliation{Physics Department E28, Technical Univerty Munich, 85748 Garching, Germany}

\author{E. Viezzer}
\affiliation{Department of Atomic, Molecular and Nuclear Physics, Universidad de Sevilla, Sevilla, 41012, Spain}
\affiliation{Centro Nacional de Aceleradores (CNA) CSIC, 41092 Seville, Spain}

\author{J. Dominguez-Palacios}
\affiliation{Department of Atomic, Molecular and Nuclear Physics, Universidad de Sevilla, Sevilla, 41012, Spain}
\affiliation{Centro Nacional de Aceleradores (CNA) CSIC, 41092 Seville, Spain}

\author{J. Rueda-Rueda}
\affiliation{Department of Atomic, Molecular and Nuclear Physics, Universidad de Sevilla, Sevilla, 41012, Spain}

\author{J. F. Rivero-Rodriguez}
\affiliation{Centro Nacional de Aceleradores (CNA) CSIC, 41092 Seville, Spain}
\affiliation{Department of Mechanical Engineering and Manufacturing, Universidad de Sevilla, Sevilla, 41092, Spain}

\author{Y. Todo}
\affiliation{National Institute for Fusion Science, 509-5292 Toki, Japan}

\author{ASDEX Upgrade team}
\altaffiliation{See author list of H. Meyer \textit{et al}. \textit{Nucl. Fusion} \textbf{59}, 112014 (2019)}
\noaffiliation

\date{\today}

\setlength{\belowcaptionskip}{-10pt}
\begin{abstract}
A synthetic Fast-Ion Loss Detector (FILD) and an imaging Heavy Ion Beam Probe (i-HIBP) have been implemented in the 3D hybrid kinetic-magnetohydrodynamic code MEGA. First synthetic measurements from these two diagnostics have been obtained for neutral beam injection (NBI) driven Alfvén Eigenmode (AE) simulated with MEGA. The synthetic fast-ion losses show a strong correlation with the AE amplitude. This correlation is observed in the phase-space, represented in coordinates $(P_\phi, E)$, being toroidal canonical momentum and energy, respectively. Fast-ion losses and the energy exchange diagrams of the confined population are connected with lines of constant $E'$, a linear combination of $E$ and $P_\phi$. First i-HIBP synthetic signals also have been computed for the simulated AE, showing displacements in the strikeline of the order of $\sim \SI{1}{mm}$, above the expected resolution in the i-HIBP scintillator of $\sim \SI{100}{\mu m}$.

\end{abstract}

\maketitle



\section{Introduction}
In magnetically confined fusion plasmas, Alfvén Eigenmodes (AEs) can be excited by various fast-ion sources, such as neutral beam injection (NBI) and fusion-born alpha particles. In turn, AEs can enhance the fast-ion transport and can lead to fast-ion losses towards the first wall in fusion devices. These uncontrolled fast-ion losses against the wall can result in hazardous heat loads in future reactors, like ITER{\cite{Heidbrink1994, Shimada2007}. 

Previous work has been carried out to characterize AEs experimentally\cite{heidbrink2018, Todo2019, KLWong1999}, focused on the identification of the fast-ion transport induced by these modes and the poloidal structures of the modes. In the ASDEX Upgrade (AUG) tokamak, the poloidal array of fast-ion loss detectors\cite{Garcia-Munoz2009,Gonzalez-Martin2018, Gonzalez-Martin2019} (FILD) have detected that the fast-ion losses are correlated with the AEs amplitude and frequency, demonstrating the AE and fast-ion interaction. Accurate characterization of these losses and the mode structures plays a key role in understanding the fast-ion confinement\cite{Garcia-Munoz2010,Garcia-Munoz2007,Garcia-Munoz_prl2010, Garcia-Munoz2019}.

In this work, these experiments are modelled with the 3D non-linear hybrid kinetic-MHD code MEGA\cite{Todo1998a}. Two synthetic diagnostics have been developed in MEGA, providing further insight into the Alfvénic activity, by studying two fundamental key points: the fast-ion loss and the radial structures.

These synthetic diagnostics are based on two of the diagnostic systems in AUG: the poloidal array of Fast-Ion Loss Detector (FILD) and the imaging Heavy-Ion Beam Probe\cite{Birkenmeier2018,Anda2018,Galdon-Quiroga2017a} (i-HIBP). For the first, a realistic 3D wall for the AUG tokamak has been implemented in MEGA showing a correlation between the fast-ion losses towards the wall and the AE activity in the simulations. For the latter, a predictive study of the synthetic signal demonstrates the capability of i-HIBP for measuring an AE located between midradius and the edge of the plasma. 

This paper is organized as follows. The model implemented in the code MEGA is briefly described in Sec. II where the implementation of the realistic 3D wall is described. The implementation of the synthetic diagnostic for i-HIBP is described in Sec. III. Sec. IV is devoted to the analysis of MEGA simulations showing the fast-ion losses and the synthetic signal for the i-HIBP diagnostic. A summary is given in Sec. V.

\section{MEGA and the realistic 3D wall}
MEGA is a numerical code that computes the self-consistent evolution of a bulk plasma and the fast-ion population in realistic 3D configurations using cylindrical coordinates. In this code, the bulk plasma is modelled using the complete non-linear single-fluid resistive-MHD equations\cite{Todo1998a,Todo2016}. Coupling between bulk plasma and fast-ion population is done via the current density in the momentum balance equation.

The set of MHD equations is spatially discretized using $4^{th}$ order finite differences on a cylindrical grid, covering the full tokamak geometry. The fast-ion distribution is sampled by markers covering the 5D reduced phase-space $(\boldsymbol{X}, p_\parallel, \mu)$, $\boldsymbol{X}$ being the guiding-center position; $p_\parallel$, parallel momentum; and $\mu$ the magnetic dipole moment. These markers are evolved using the gyro-kinetic equations with Finite Larmor Radius (FLR) corrections. The fast-ion distribution is evolved using \textit{particle-in-cell} with the $\delta f$ method\cite{Dimits1993}.

The cylindrical grid resolution has been chosen to be $(N_R, N_\phi, N_z) = (128, 64, 256)$, allowing the evaluation up to the $n=5$ toroidal mode number. Higher toroidal mode numbers are filtered out to reduce numerical noise. The time evolution is obtained using an explicit $4^{th}$ order Runge-Kutta scheme.

\begin{figure}
    \centering
    \includegraphics[width=6.0cm]{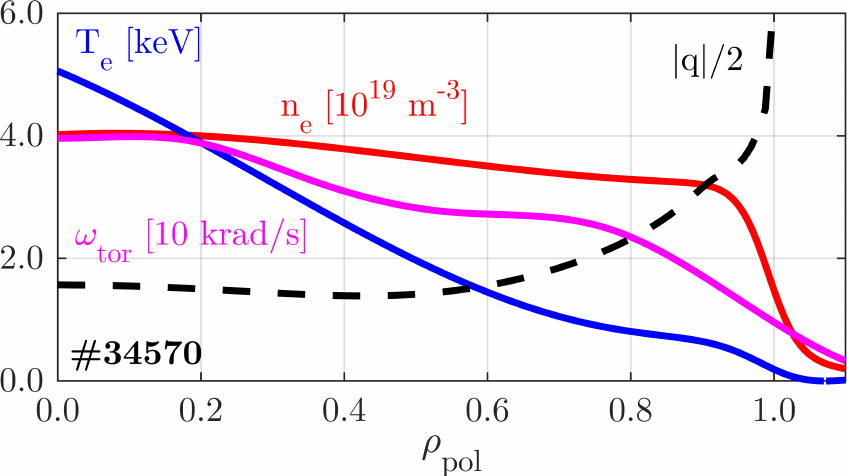}
    \caption{Density, temperature and toroidal rotation used as starting point for MEGA simulations, taken from the AUG tokamak discharge $\#34570$ ($t = \SI{3.53}{s}$). The reconstructed $q$-profile is shown in dashed lines.}
    \label{fig:equilibrium}
\end{figure}

Experimental profiles and magnetic reconstructions from AUG pulse $\#34570$ are used as inputs for the MEGA simulations. In Fig. \ref{fig:equilibrium} the initial profiles are presented, corresponding to a discharge at the AUG tokamak. The initial fast-ion distribution is given by a modelled off-axis NBI slowing-down distribution using a Gaussian term\cite{Gonzalez-Martin2019} for the spatial dependence. The initial distribution function used in this work is:

\begin{multline}
    \label{eq:spt_part_fidist}
    F_\text{phase-space} \propto e^{-\frac{\left(\rho-\rho_0\right)^2}{2\left(\Delta\rho\right)^2}}\frac{1}{v^3+v^3_\text{crit}}\text{erfc}\left(\frac{v-v_\text{birth}}{\Delta v}\right)\times\\ e^{-\frac{\left(\Lambda-\Lambda_0\right)^2}{2\left(\Delta\Lambda_0\right)^2}}
\end{multline}
being $\rho$ the normalized poloidal magnetic flux. Spatial parameters have been fixed to $\rho_0 = 0.4$, $\Delta\rho= 0.15$ in this work. $v_\text{birth}$ is the birth velocity, has been set to $\SI{93}{keV}$, and $\Delta v = 0.05\cdot v_\text{birth}$. For the pitch-angle, a Gaussian dependence in $\Lambda \equiv 1-\lambda^2 = \frac{\mu B}{E}$ is introduced. The pitch-angle parameters has been set to $\Lambda_0 = 0.55$ and $\Delta\Lambda_0 = 0.20$ in this work. Finally, $v_\text{crit}$ is the critical velocity\cite{Jacquinot1999}.

The parameter regulating the fast-ion density, $\beta_\text{fi}=\frac{p_\text{fi}}{B_\text{axis}^2/2\mu_0}$, is set to $\beta_\text{fi} = 0.67\%$, being $p_\text{fi}$ the fast-ion pressure. This value corresponds to the NBI6 for the same discharge. Note that this work does not intend to reproduce the Toroidal Alfvén Eigenmodes (TAEs) observed in the experiment, but to provide a suitable simulation case to test the newly implemented synthetic diagnostics. A detailed comparison to the experiment is ongoing\cite{Gonzalez-Martin2021}.

\begin{figure}[h!]
    \centering
    \includegraphics[width=3.7cm]{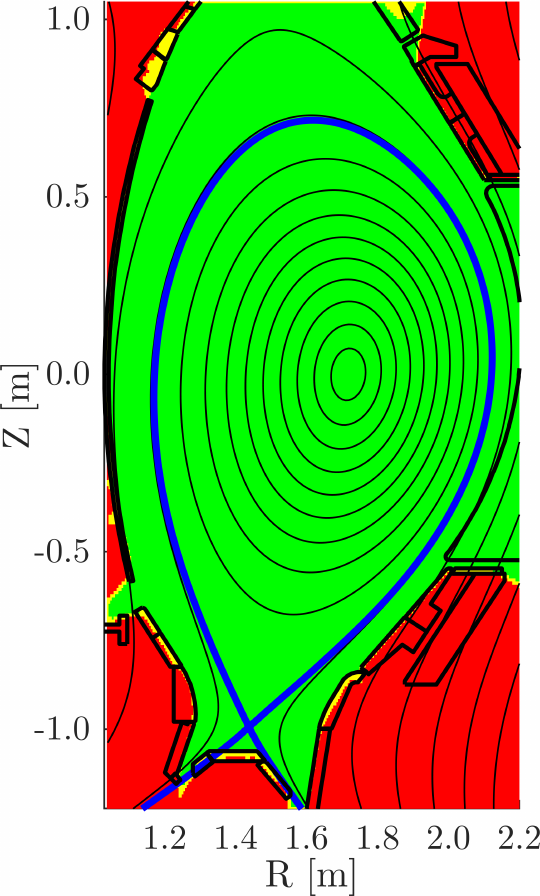}
    \caption{Map of the wall (here represented for $\phi = 0^o$) used in MEGA to obtain the fast-ion losses. Green and red regions indicate inside/outside simulation domain for the FI markers. The first wall is used as limit where fast-ion markers are captured. Solid lines represent flux surfaces. In blue, the separatrix is indicated.}
    \label{fig:wall}
\end{figure}

A 3D wall has been implemented in MEGA, allowing to stop the evolution of fast-ion markers when they reach the first wall. The mapping of the wall to the cylindrical grid (where the MHD equations are solved) allows for a fast implementation of the 3D wall, without a significant impact on the simulation efficiency (less than $1\%$). In Fig. \ref{fig:wall}, a poloidal cut of the mapped wall in MEGA is shown. The regions marked in yellow (some of them lie behind the 2D wall model represented by the thicker black lines) determine where the fast-ion evolution will be stopped and considered as fast-ion losses (FIL).

\section{Synthetic i-HIBP diagnostic}

The i-HIBP diagnostic injects a heavy-neutral primary beam (${}^{133} \text{Cs}$ or ${}^{85,87} \text{Rb}$) into the plasma that ionizes due to multiple processes. These ionized particles, forming the secondary beam, start a gyromotion until reaching a scintillator plate. The signal on the scintillator translates into a two-fold information: the intensity of the strike line provides the plasma density ($n_e$); and the strike line position and shape provides information on the magnetic and electric fields, $\boldsymbol{B}$ and $\boldsymbol{E}$, respectively\cite{Galdon-Quiroga2017a}.

The synthetic diagnostic of i-HIBP, the new \textit{i-HIBPsim} code, is based on kinetic simulations for the two main species, the heavy-neutrals (primary beam) and heavy-ions (secondary beam). Markers are launched at the injection port and tracked into the plasma using a Boris leap-frog scheme\cite{Boris1971}. In this work, an infinitely small beam is used, i.e., the width of the beam and divergences are set to zero. The secondary beam birth distribution is obtained by using a beam attenuation model:
\begin{equation}
    \label{eq:beam_attenuation_model}
    \dot{W}_j = - \sum_{k\in\text{reactions}} W_j n_k \langle\sigma v\rangle_k,
\end{equation}
where the sum is over all possible reactions that attenuate the beam; $n_k$ is the secondary reactant density; and $\langle\sigma v\rangle_k$ is the reaction rate of the $k^\text{th}$ reaction. Only two reactions are considered to generate the attenuation of the primary: the electron-impact ionization\cite{Lotz1967} and the charge-exchange reactions with main-ions\cite{Meyer1975,Ebel1987,Girnius2002}. Single-ionization step is implemented in the simulation code as the recombination via charge-exchange (i.e., $Cs^++D^0\rightarrow Cs^0+D^+$) is expected to be much smaller, since the neutral density in AUG is of the order\cite{Viezzer2011} of $n_0 \sim 10^{16}\ \text{m}^{-3}$. Impurities induced ionizaation reaction rates, as extrapolated from lithium in \cite{Wutte1997}, are negligible compared to the main ion charge-exchange and electron-impact ionization rates. This, combined with the typical impurity concentration ($\sim 1\%$ after the boronization\cite{Ryter2013}), makes this interaction negligible, compared collisions with electrons and main ions.

The secondary beam will travel following gyroorbits until hitting the scintillator, determined via a ray-triangle algorithm\cite{Moller1997}.
The beam-attenuation equation is also used to determine the secondary beam flux into the scintillator. A single-step ionization is used for the secondary beam considering only the electron-impact ionization\cite{Hertling1982, Higgins1989}. Charge-exchange recombination for the secondary beam is not taken into account, as for the primary beam.

Markers evolve in a fully 3D input electromagnetic field, allowing for a direct connection with MEGA. The electromagnetic perturbations computed by MEGA for a certain plasma phenomena, can be used to feed the synthetic diagnostic and obtain the predicted signal. 

In previous experiments in fusion devices, like TJ-II, a similar diagnostic, the HIBP, has been used to detect and characterize the poloidal mode numbers and structure of AE\cite{Jimenez-Gomez2011}. The scintillator-based i-HIBP will provide the high spatial resolution measurements, as shown in Fig. \ref{fig:rbstrikemap} (up to $~\SI{100}{\mu m}$ in the scintillator). 

\section{Simulation results}
The off-axis fast-ion distribution produces in the plasma a TAE located at the midradius ($\rho_\text{pol} = 0.70$). This example case, using the realistic 3D wall, already suggests an important fast-ion loss mechanism, via the wave-particle resonances. The location of the AE, as shown in Fig. \ref{fig:energyevol}(b), is close enough to the edge for the heavy-ions in i-HIBP to feel the perturbations. The green lines represent trajectories for ${}^{85} \mathrm{Rb}$ under the perturbed magnetic field, proving the possible range of detection.

\begin{figure}
    \centering
    \includegraphics[width = 5.7cm]{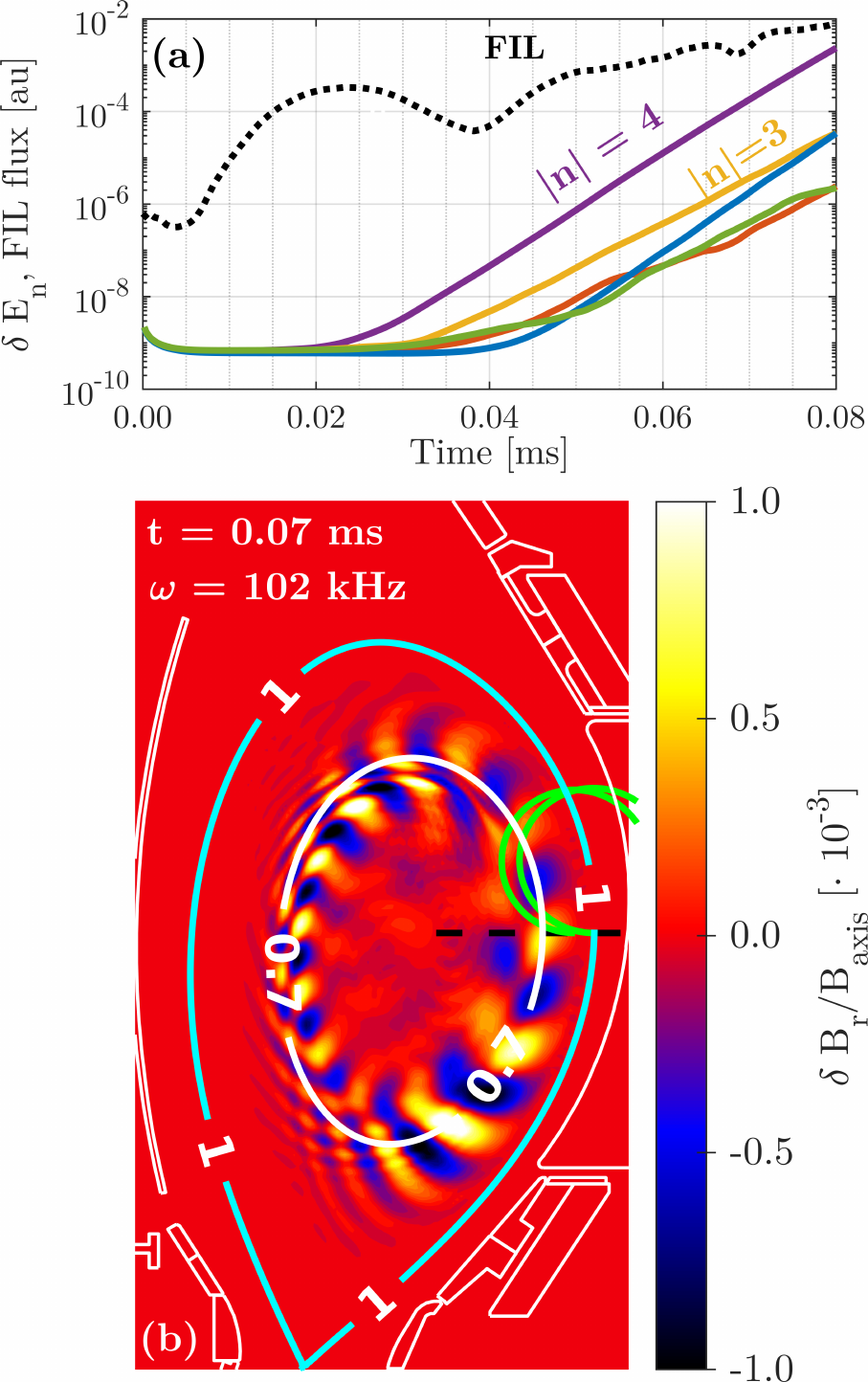}
    \caption{(a) In solid lines, evolution of the energy associated to the toroidal $\left|n\right|$ mode numbers. In dashed lines, the FIL flux associated to the region in velocity-space $\mu\in\left(3.0, 5.0\right)\cdot 10^{-15}\rm{ J/T}$. (b) Poloidal representation of the perturbation at $t = \SI{0.07}{ms}$. Flux surfaces corresponding to $\rho_\text{pol} = 0.70, 1.00$ as well as the 2D wall structures, have been represented for visual guidance. In green, ${}^{85} \mathrm{Rb}$ orbits obtained with \textit{i-HIBPsim} and discussed in Section \ref{subsec:synth_ihibp}. }
    \label{fig:energyevol}
\end{figure}

The implementation of an off-axis initial fast-ion spatial distribution leads to Alfvénic activity closer to the plasma edge. In Fig. \ref{fig:energyevol}(a), the toroidal mode energy in the bulk plasma is shown on a logarithmic scale. The $\left|n\right|=4$ mode shows the largest growth and dominates the plasma phenomena. In dashed, the fast-ion \textit{flux} for given $\mu\in\left(3.0, 5.0\right)\cdot 10^{-15}\ \text{J/T}$ interval has been represented. Note that to avoid the inclusion of unrealistic fast-ion prompt losses in the simulation the quantity $\left|\delta w_j\right| N_j$ has been presented instead, where $\left|\delta w_j\right|$ is the differential weight evolution with respect to the equilibrium, and $N_j$ is the number of particles represented initially by the marker. The markers strongly interacting with the mode have a higher $\left|\delta w_j\right|$, hence allowing us to focus on fast-ion loss induced by the mode.

In Fig. \ref{fig:energyevol}(b) the poloidal structure of the mode at $t = \SI{0.07}{ms}$ is shown. The mode is located around the surface $\rho_\text{pol} = 0.7$, superimposed for visual guidance. A Fourier transform of its time evolution shows that the frequency of the mode is $f = \SI{102}{kHz}$.

\subsection{Synthetic fast-ion losses}

In the AUG tokamak, strong TAE-coherent fast-ion losses have been detected with the fast-ion loss detector poloidal array\cite{Garcia-Munoz_prl2010}. The induced losses can be explained by the magnetic perturbation producing an open trajectory, without a net energy exchange; or via a power exchange with the mode resulting in an orbit kick away from the confined region.
 
Fast-ion losses in MEGA simulations have been obtained using a self-consistent approach, capturing the fast-ion markers during the simulations. To identify whether these losses are produced by a significant interaction with the AE, the phase-space, represented in variables $\left(P_\phi,E,\mu\right)$, being toroidal canonical momentum, energy and magnetic moment, respectively. In Fig. \ref{fig:filanalysis} (a-b) the phase-space for $\mu\in\left(3.0, 5.0\right)\cdot 10^{-15}\rm{J/T}$ is shown. For the confined population (a) the instantaneous energy exchange is presented, indicating the region where the mode is interacting the strongest with the fast-ions. For the FIL (b) the impinging flux onto the 3D wall is presented.

\begin{figure}
    \centering
    \includegraphics[width=6.3cm]{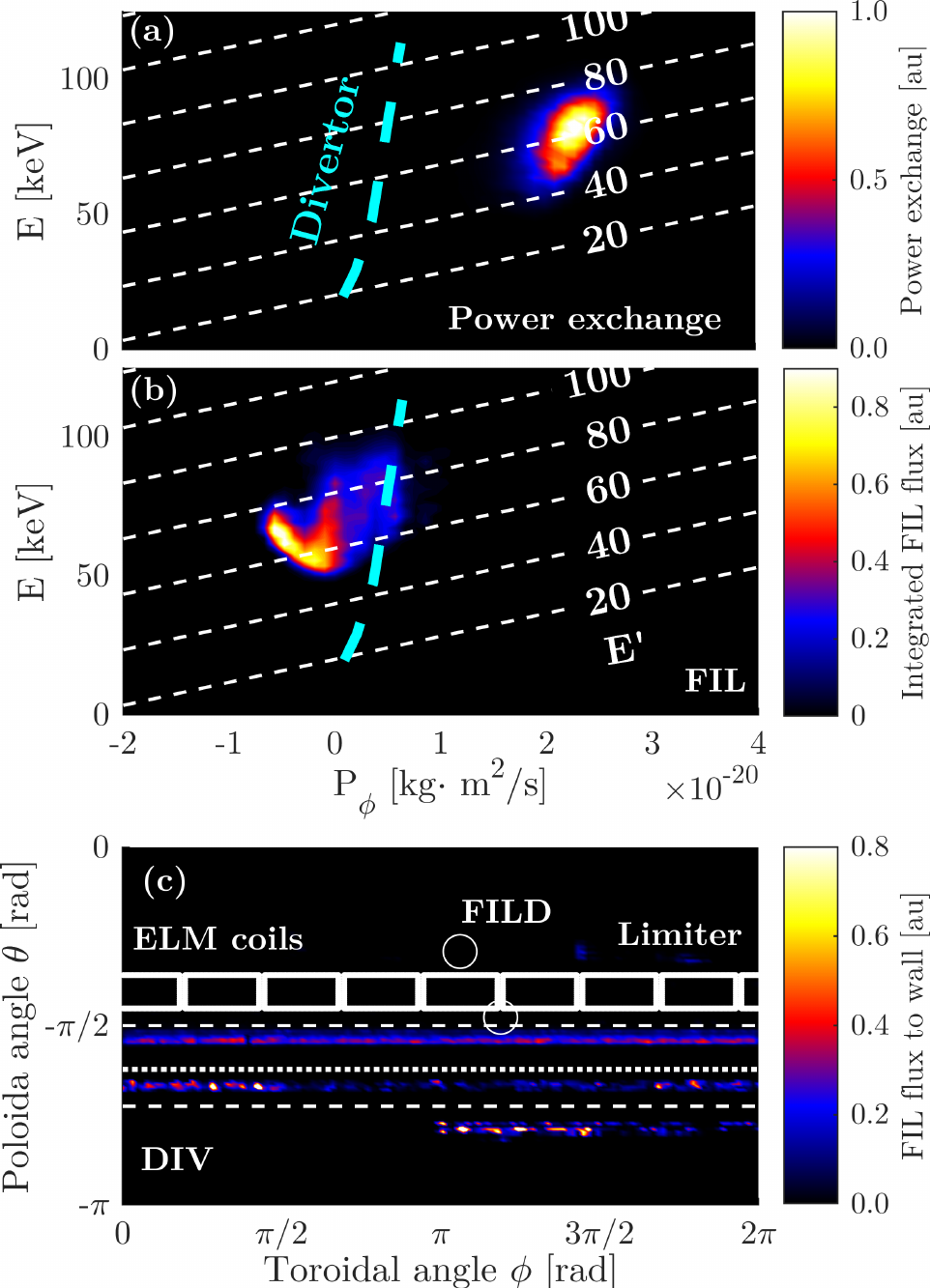}
    \caption{(a-b) Slice of the velocity space for the fast-ion, corresponding to $\mu\in\left(3.0,5.0\right)\cdot 10^{-15}\rm{ J/T}$. (a) Instantaneous power exchange of the confined fast-ion population at $t = \SI{0.059}{ms}$, superimposed with the lines of constant $E'$. Dashed blue lines represents the divertor surface projected on the velocity space. (b) Fast-ion loss flux (number of fast-ion hitting the wall) for times $t > 0.060\rm{ ms}$. (c) Toroidal plane of the fast-ion losses $t>0.060\rm{ ms}$, where important structures have been represented. Here $\theta=0$ represents the midplane and $\theta = -\pi$, the lower divertor.}
    \label{fig:filanalysis}
\end{figure}

In the presence of a wave with a given low and constant frequency, $\omega_n$, and toroidal mode number $n$, the conserved quantity\cite{Todo2019} is $E' = E - \frac{\omega_n}{n}P_\phi$. Contour lines with this quantity have been superimposed in both Fig. \ref{fig:filanalysis}(a,b). Fast-ions that drive the AE, drift away and hit the wall following constant $E'$ lines. As a visual guidance, the divertor region is represented in both figures as cyan dashed line, which is the region with the largest heat loads.  In Fig. \ref{fig:filanalysis}(c), the FIL flux is represented in the angular plane only below the midplane ($\theta = 0$), since it is the region receiving most of the FIL flux.

A simulation with only the toroidal mode numbers $\left|n\right|=0, 4$, i.e., filtering the rest of the Fourier components, shows that both the confined population energy exchange is still present in the same location. The FIL in these simulations do not differ significantly from the multi-n simulation. We can conclude that the fast-ion transport and losses are dominated by their interaction with the toroidal mode number $n = 4$. 

This analysis suggests that the fast-ion losses are predominately expulsed from the plasma due to the energy exchange with the AE during the linear growth phase. This would play a central role in the interpretation of the FILD experiments in the AUG tokamak, since it connects the fast-ion loss measurements to an energy exchange with the mode.

\subsection{Synthetic i-HIBP signal}
\label{subsec:synth_ihibp}
The simulated AE is used here to construct the i-HIBP synthetic signal. The radial structures, as shown in Fig. \ref{fig:energyevol}(b) are located in an off-axis location, $\rho_\text{pol}\approx 0.55-0.75$.  The total mode amplitude is of the order of $\delta B_r/B_\text{axis} \sim 10^{-3}$, typical from NBI-driven AE experiments in the AUG tokamak, allowing us to obtain realistic estimates of the i-HIBP signal.

\begin{figure}
    \centering
    \includegraphics[width=4.6cm]{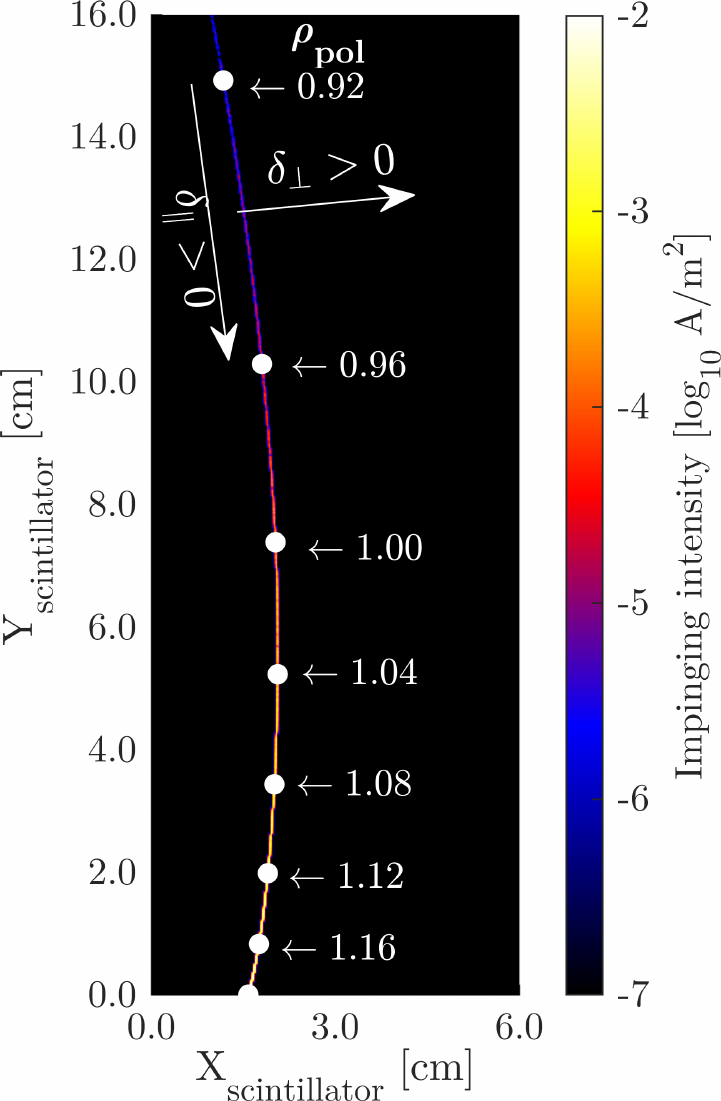}
    \caption{Strikeline representation for the equilibrium case for the ${}^{87}\mathrm{Rb}$ case. In white dots, the origin points of the secondary has been displayed. The convention for strike line perturbations is defined  in the figure: perpendicular to the strike line (mostly in the X-direction) is positive to the right.}
    \label{fig:rbstrikemap}
\end{figure}

The signal on the scintillator has been obtained by tracking both the primary and secondary beam using the \textit{i-HIBPsim} synthetic diagnostic, described in Section III. The long gyro-radius (of the order of $\SI{20}{cm}$ in the current setup, with $E = \SI{70}{keV}$ for the ${}^{85,87}\mathrm{Rb}$, $E=\SI{50}{keV}$ for the ${}^{133}\mathrm{Cs}$; and $B = \SI{2.5}{T}$) of the heavy-ion takes them further within the plasma where electromagnetic perturbations modify their orbits before reaching the scintillator. In Fig \ref{fig:rbstrikemap}, the strikeline on the scintillator for the ${}^{87}\mathrm{Rb}$ is shown for the baseline scenario, i.e., without perturbations. For visual guidance, points with the birth $\rho_\text{pol}$ location are superimposed, and the sign convention for the line perturbation is presented.

Simulations with and without the AE perturbations have been carried out in order to assess the changes in the strike line on the scintillator. The perturbation case has been analyzed with and without the generated AE electric field ($\delta E_r \sim \SI{6}{kV/m}$), in order to isolate the impact of the electric field on the strike pattern. Two key parameters are studied to determine the impact on the strike line: the perpendicular displacement of the strikeline (perpendicular to the case without perturbation), $\delta_\bot$ in Fig. \ref{fig:ihibpdisp}(a); the modification of the intensity in the strike line due to the density and temperature perturbations, in Fig. \ref{fig:ihibpdisp}(b). Both species, ${}^{85,87}\mathrm{Rb}$ and ${}^{133}\mathrm{Cs}$, available as separate sources in the AUG i-HIBP diagnostic, have been used in the analysis.

\begin{figure}
    \centering
    \includegraphics[width=6.7cm]{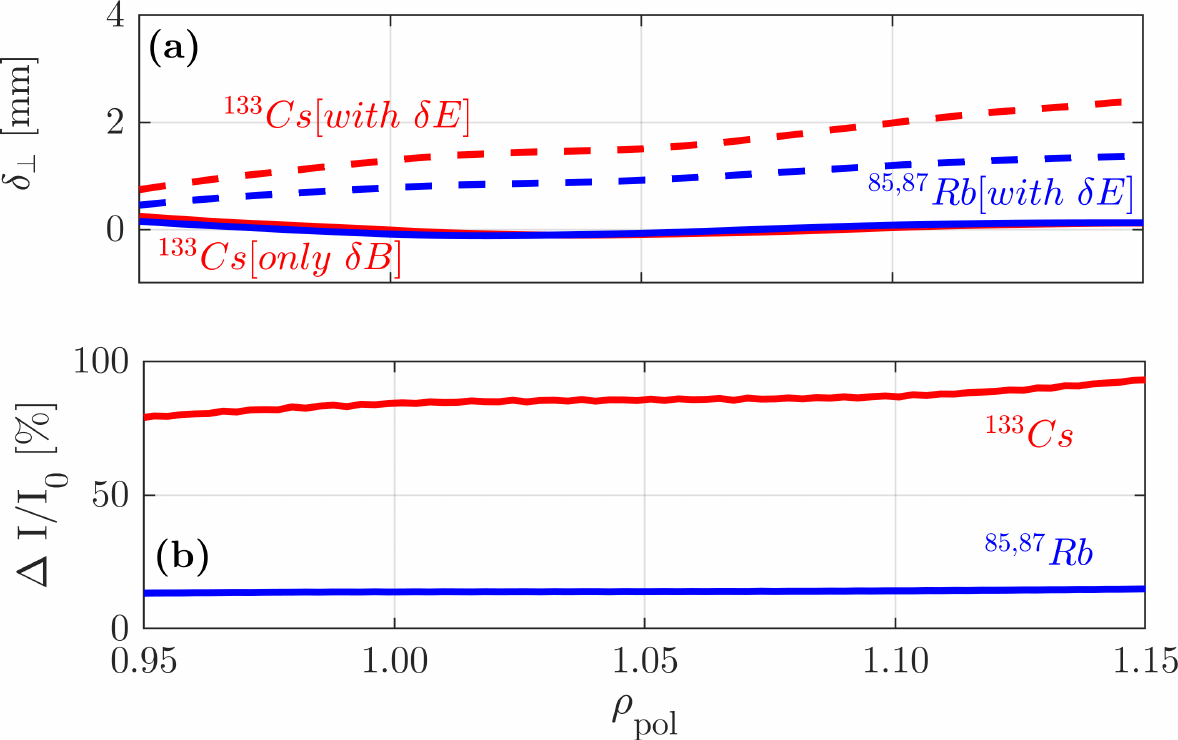}
    \caption{Impact on the strike line of the simulated TAE. (a) Perpendicular displacement (with respect to the one defined in Fig. \ref{fig:rbstrikemap}) of the strikeline due to magnetic perturbation alone (solid lines) and including both electric and magnetic perturbations (dashed lines). (b) Relative deviation in the intensity seen in the strikeline with respect to the equilibrium. Dashed and solid lines are superimposed in this case. For ${}^{133}\mathrm{Cs}$ case, relative variation reaches up to $\sim 75\%$.}
    \label{fig:ihibpdisp}
\end{figure}
The relevant comparison in the strikeline is the case without electric field (solid lines) and with the electric field (dashed lines). For both species, the displacement caused only by the magnetic field perturbation lies in the range of $\sim \SI{40}{\mu m}$, while the electrostatic potential induces a deviation up to $\sim \SI{2}{mm}$, above the expected optical resolution ($\sim \SI{100}{\mu m}$) on the scintillator. For the intensity pattern variation, Fig. \ref{fig:ihibpdisp}(b), the deviation due to the electric perturbation is negligible. The relative pattern variation is an order of magnitude different between $\mathrm{Cs}$ and $\mathrm{Rb}$ beams, being $\approx 75\%$ and $\approx 15\%$ respectively. This systematic deviation in intensity points to the possibility of measuring $\delta n_e$ due to Alfvénic modes in i-HIBP.

\section{Conclusions}
In this work, the synthetic diagnostics for fast-ion losses and the i-HIBP diagnostic have been developed and applied to a MEGA simulation. The synthetic fast-ion loss diagnostic shows already promising results, connecting the fast-ion losses to the interaction with Alfvénic phenomena, through $E'$ lines. This methodology can now be extended to the rigorous study of plasma instabilities and understanding the fast-ion losses associated to them. 

The preliminary study of the synthetic i-HIBP signal for the simulated AE shows that the radial structures may be resolved by the i-HIBP diagnostic. The impact on the strikeline due to the electric perturbation induced by the AEs will be measurable with the i-HIBP scintillator ($\approx \SI{2}{mm}$).

\section*{Acknowledgements and data availability}
This work received funding from the European Starting Grant (ERC) from project 3D-FIREFLUC and from the Spanish Ministry of Science under Grant No. FPU19/02267. This work has been carried out within the framework of the EUROfusion Consortium and has received funding from the Euratom research and training programme 2014-2018 and 2019-2020 under grant agreement No 633053. The views and opinions expressed herein do not necessarily reflect those of the European Commission

\section*{Data availability statement}
The data that support the findings of this study are available from the corresponding author upon reasonable request.

\section*{References}

\bibliography{HTPD_Oyola}

\end{document}